\documentclass[12pt]{article}

\usepackage{geometry}
\usepackage{authblk}
\usepackage{graphicx}%
\usepackage{multirow}%
\usepackage{amsmath,amssymb,amsfonts}%
\usepackage{amsthm}%
\usepackage{mathrsfs}%
\usepackage[title]{appendix}%
\usepackage{xcolor}%
\usepackage{textcomp}%
\usepackage{manyfoot}%
\usepackage{booktabs}%
\usepackage{algorithm}%
\usepackage{algorithmicx}%
\usepackage{algpseudocode}%
\usepackage{listings}%
\usepackage{hyperref}
\usepackage{subcaption}
\hypersetup{hidelinks,
	colorlinks=true,
	allcolors=black,
	pdfstartview=Fit,
	breaklinks=true}
\usepackage{makecell} 

\newcommand{\ours}{\textbf{\texttt{MLM-FG}}}

\title{Pre-trained Molecular Language Models with Random Functional Group Masking\footnote{Corresponding to Haoyi Xiong via haoyi.xiong.fr@ieee.org}}
\begin{document}

\author[1,2,$\dag$]{Tianhao Peng} 
\author[1,3]{Yuchen Li}
\author[1]{Xuhong Li}
\author[1]{Jiang Bian}
\author[4]{Zeke Xie}
\author[5]{Ning Sui}
\author[6,7]{Shahid Mumtaz}
\author[8]{Yanwu Xu}
\author[3]{Linghe Kong}
\author[1,$\dag$]{Haoyi Xiong}

\affil[1]{Baidu Research, Baidu Inc., Haidian, 100085, Beijing, China.}
\affil[2]{School of Computer Science and Engineering, Beihang University, Haidian, 100091, Beijing, China.}
\affil[3]{Department of Computer Science and Engineering, Shanghai Jiao Tong University, Minhang, 200240, Shanghai, China.}
\affil[4]{The Hong Kong University of Science and Technology (Guangzhou), Guangzhou, 511455, Guangdong, China.}
\affil[5]{Department of Molecular and Structural Biochemistry, North Carolina State University, Raleigh, NC 27695, United States}
\affil[6]{Department of Computer Science, Nottingham Trent University, Nottingham, UK.}
\affil[7]{Department of Applied Informatics, Silesian University of Technology, Gliwice, Poland.}
\affil[8]{School of Future Technology, South China University of Technology, Guangzhou, 510641, Guangdong, China.}
\affil[$\dag$]{These authors made equal technical contribution to this work.}

\date{}
\maketitle
\abstract{Recent advancements in computational chemistry have leveraged the power of trans-former-based language  models, such as MoLFormer, pre-trained using a vast amount of simplified molecular-input line-entry system (SMILES) sequences, to understand and predict molecular properties and activities, a critical step in fields like drug discovery and materials science. To further improve performance, researchers have introduced graph neural networks with graph-based molecular representations, such as GEM, incorporating the topology, geometry, 2D or even 3D structures of molecules into pre-training. While most of molecular graphs in existing studies were automatically converted from SMILES sequences, it is to assume that transformer-based language models might be able to implicitly learn structure-aware representations from SMILES sequences.
In this paper, we propose \ours{} -- a SMILES-based \underline{\em M}olecular \underline{\em L}anguage \underline{\em M}odel, which randomly masking SMILES subsequences corresponding to specific molecular \underline{\em F}unctional \underline{\em G}roups to incorporate structure information of atoms during the pre-training phase. This technique aims to compel the model to better infer molecular structures and properties, thus enhancing its predictive capabilities.
Extensive experimental evaluations across 11 benchmark classification and regression tasks in the chemical domain demonstrate the robustness and superiority of \ours{}. Our findings reveal that \ours{} outperforms existing pre-training models, either based on SMILES or graphs, in 9 out of the 11 downstream tasks, ranking as a close second in the remaining ones. Remarkably, \ours{} also surpasses 3D-graph-based models, which explicitly incorporate molecular structures into their inputs, highlighting its exceptional capacity for representation learning even without explicit 3D structural information. These results indicate that \ours{} effectively captures the nuanced language of molecules, offering promising implications for computational chemistry and related disciplines.}



\section{Main}
While deep learning has been widely explored in cheminformatics with significant progress~\cite{huang2016communication, DBLP:journals/jcheminf/DavidTME20}, her potential is severely limited by the scale of labeled data.
%
%
To reduce the cost of data annotation while enabling generalizable, transferable, and robust representation learning from unlabeled data, researchers extend the pre-training strategies~\cite{DBLP:conf/naacl/DevlinCLT19} from images and texts to molecular data~\cite{DBLP:journals/natmi/WangWCF22, DBLP:journals/corr/abs-2106-10234, DBLP:journals/corr/abs-2012-11175, DBLP:conf/nips/RongBXX0HH20, ross2022large}. 
%
%


To work with machine learning algorithms, 
molecules can be represented by a chemical notation language named SMILES~\cite{weininger1988smiles}, which explicitly represents meaningful substructures such as branches and cyclic structures. To enable pre-training, researchers use the variant of Transformers to pre-train on large-scale unlabeled SMILES strings~\cite{wang2019smiles, ross2022large,broberg2022pre}. 
To pre-train a molecular language model, given the SMILES string of every molecule in the training dataset, existing methods usually adopt an masked autoencoding strategy~\cite{ross2022large,wang2019smiles,broberg2022pre, irwin2022chemformer}. It first randomly selects a subsequence of the SMILES string. Then the strategy masks the selected part and trains models to predict the masked part. However, such random masking strategy would ignore the key chemical substructures of molecules, such as rings and functional groups~\cite{DBLP:conf/nips/RongBXX0HH20, DBLP:journals/natmi/WangWCF22}. 
For instance, consider aspirin, which is denoted by ``O=C(C)Oc1ccccc1C(=O)O''. In this molecular structure, critical functional groups such as the carboxylic acid (``-COOH'') and the ester (``-COO-'') are at risk of being overlooked due to random masking. This oversight neglects their pivotal contributions to the molecular activity and properties.
To the end, these methods may fail to \emph{learn the critical molecular properties, which are primarily relevant to the chemical substructures of a molecule, from SMILES strings}~\cite{DBLP:conf/nips/ZhangLWLL21, DBLP:journals/corr/abs-2106-04509}. 


The previous investigation has highlighted the limitations of SMILES in terms of topology awareness, underscoring its inability to explicitly encode structural information of molecules\cite{zhang2023applications}. To address the issue, structure-aware pre-training methods utilizing Graph Neural Networks (GNNs) have marked a significant advancement. These approaches leverage the graph-format representation of molecules (such as the topology of every atom in a molecule onto a 2D space), enriching the learning models with a deeper understanding of molecular structures~\cite{DBLP:conf/nips/RongBXX0HH20, DBLP:journals/corr/abs-2012-11175, DBLP:journals/natmi/WangWCF22}. More recently, 3D graph-based molecular data representation has been introduced in GNN pre-training, where the structures of massive molecules in a 3D space have been used to boost the performance of pre-trained models by incorporating the 3D structural/topological information~\cite{fang2022geometry, atz2021geometric}. However, such structural/topological information is not also precise. For example, in addition to acquiring the 3D positions of every atom and the angles between every two bonds in a molecule through experiments, some studies directly convert SMILES strings or 2D topology graph of a molecule into a 3D graph using Merck molecular force field (MMFF94)~\cite{halgren1996merck} function in RDKit~\cite{fang2022geometry}. It is reasonable to doubt the ``add-value'' of such automatic data format conversion and the precision of converted 3D graphs. Thus, it would be challenging to \emph{pre-train a structure-aware molecular model while the precise structural information is not available.}

\begin{figure*}
	\centering 
	\includegraphics[width=0.95\linewidth]{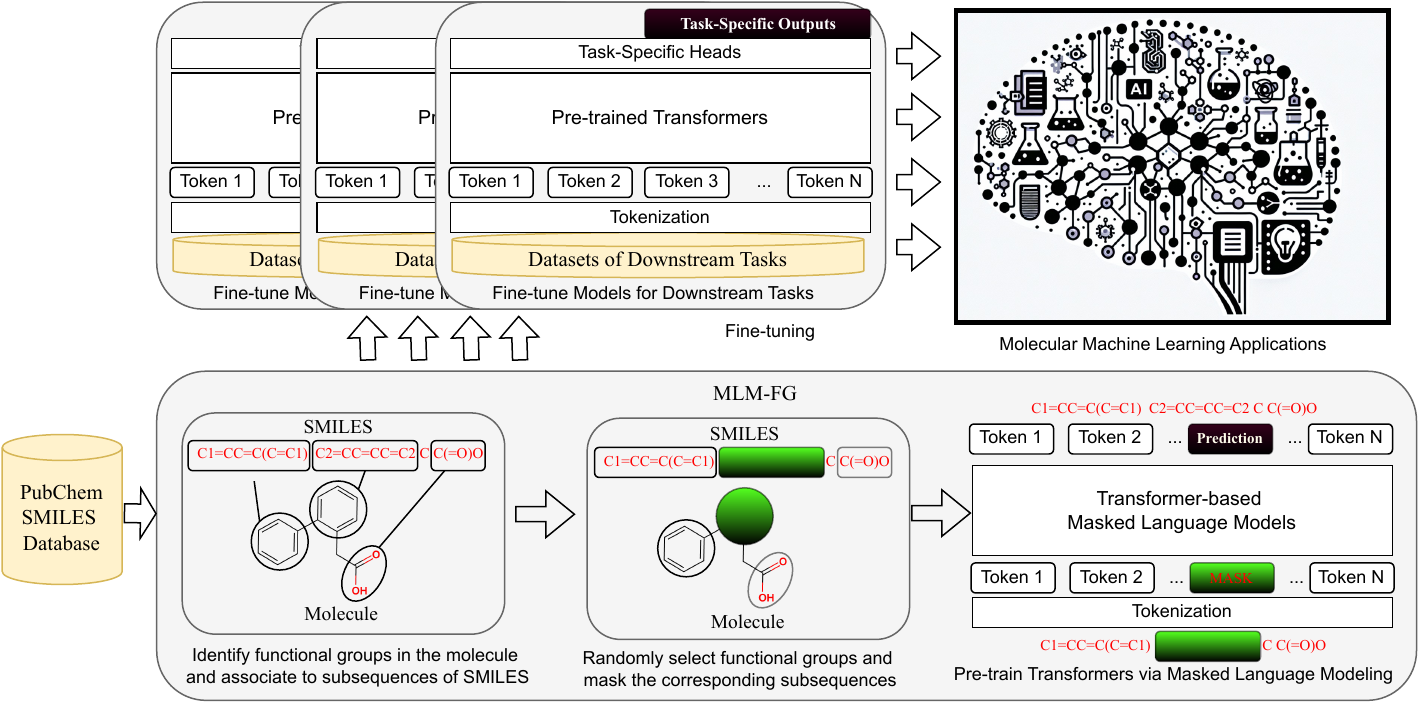}
	\caption{An illustration of the proposed \ours{} framework: (1) \ours{} adopts 12-layer multi-head transformer blocks (in either RoBERTa or MoLFormer architectures) with a hidden state dimension of $D_h$=768 for pre-training and fine-tuning, (2) \ours{} follows a functional group-aware random masking strategy to pre-train the model on a large corpus of 10 to 100 million SMILES sequences from PubChem, and (3) \ours{} fine-tunes the pre-trained models to support a wide range of molecular machine learning applications.}
    \label{fig:framework}
\end{figure*}

To tackle the aforementioned challenges, we propose a novel molecular representation framework \ours{} -- a SMILES-based \underline{\em M}olecular \underline{\em L}anguage \underline{\em M}odel, which randomly masking SMILES subsequences corresponding to specific molecular \underline{\em F}unctional \underline{\em G}roups to incorporate structure information of atoms during the pre-training phase. 
Specifically, \ours{} employs transformer-based models trained on a large corpus of SMILES strings for 100 million molecules. As shown in Figure~\ref{fig:framework}, given the SMILES string for every molecule in the training dataset, \ours{} first parses the string and identify the subsequences corresponding to functional groups and key clusters of atoms in the molecules. Then \ours{} randomly masks a certain proportion of subsequences and trains the model to predict the masked part as the pre-training task. 

Extensive experimental evaluations across 11 benchmark classification and regression tasks in the chemical domain demonstrate the robustness and superiority of \ours{}. Our findings reveal that \ours{} outperforms existing pre-training models, either based on SMILES or graphs, in 9 out of the 11 downstream tasks, ranking as a close second in the remaining ones. Remarkably, \ours{} also surpasses 3D graph-based models, which explicitly incorporate molecular structures into their inputs, highlighting its exceptional capacity for representation learning even without explicit 3D structural information. These results show that pre-trained transformer encoders specialized in molecular SMILES demonstrate robust performance, matching or even exceeding existing supervised or unsupervised language models and GNN benchmarks in accurately forecasting a broad spectrum of molecular properties. 

\section{Results}
In this section, we present a series of comprehensive experiments designed to illustrate the efficacy of \ours{}. These experiments include performance comparisons across various downstream tasks and visual analysis of pre-trained representations. To assess the impact of model architecture and data size, we utilized two transformer-based models for pre-training on a corpus consisting of millions of molecules. These models are based on the MoLFormer~\cite{ross2022large} and RoBERTa architectures. 

Moreover, we conducted a comparative analysis of \ours{} with models pre-trained using different strategies and methods derived from existing literature. These include models based on both molecular graphs, such as MolCLR~\cite{DBLP:journals/natmi/WangWCF22}, GROVER~\cite{rong2020grover}, and GEM~\cite{fang2022geometry}, and SMILES e.g., MoLFormer~\cite{ross2022large}. Notably, two recent works -- GEM~\cite{fang2022geometry}, incorporating the explicit 3D structures of 20 million molecules in pre-training, and MoLFormer~\cite{ross2022large}, pre-trained using SMILES strings of 1.1 billion molecules, are two strong baselines in the line of research for molecular graph-based and SMILE-based solutions.

\subsection{Performance of \ours{} on Downstream Tasks}
Before fine-tuning \ours{} to downstream tasks, we use 10 million, 20 million, and 100 million unlabelled molecules sampled from PubChem~\cite{kim2019pubchem}, a public access database that contains purchasable drug-like compounds, to pre-train \ours{} on two transformer-based models. Subsequently, we conduct experiments on multiple molecular benchmarks from the MoleculeNet~\cite{wu2018moleculenet}, including seven classification tasks and five regression tasks. Following the previous work~\cite{fang2022geometry,hu2020strategies,ross2022large}, we adopt the scaffold split~\cite{ramsundar2019deep} to split the datasets, which splits molecules based on their molecular substructure. 
By separating structurally distinct molecules into different subsets, scaffold splitting poses a more significant challenge and offers a robust test of model generalizability compared to random splitting methods. 

\begin{table*}[!t]
\centering
\caption{Comparison of classification accuracy between fine-tuned \ours{} and existing pre-trained/self-supervised baselines on multiple classification benchmarks.}
\scalebox{0.8}{
\begin{tabular}{cccccccc}
\toprule
& BBBP   & BACE   & ClinTox & Tox21  & SIDER  & HIV    & MUV    \\ \hline
No. molecules & 2,039 & 1,513 & 1,478 & 7,831 & 1,427 & 41,127 & 93,087 \\
No. prediction tasks & 1 & 1 & 2 & 12 & 27 & 1 & 17 \\
\hline
\multicolumn{8}{c}{Pre-trained models from existing literature}\\ \hline
MolCLR-gin                 & \underline{0.9307} & 0.7873 & 0.8005  & 0.7644 & 0.5826 & 0.7768 & 0.7386 \\
MolCLR-gcn                 & 0.8432 & 0.7194 & 0.7997  & 0.7179 & 0.5353 & 0.7616 & 0.6701 \\
GROVER-base                & 0.9022 & 0.7700 & 0.6847  & 0.7187 & 0.5579 & 0.6950 & 0.6265 \\
GROVER-large               & 0.8861 & 0.7795 & 0.6082  & 0.7155 & 0.5283 & 0.6956 & 0.5132 \\
GEM                        & 0.9103 & \textbf{0.8603} & 0.8506  & 0.7791 & \textbf{0.6279} & 0.7500 & 0.7253 \\
MoLFormer                  & 0.9037 & 0.8275 & 0.9451  & 0.7734 & 0.5826 & 0.7630 & \underline{0.7599} \\ \hline
\multicolumn{8}{c}{MoLFormer and RoBERTa models without pre-training}\\ \hline
MoLFormer (from scratch) & 0.8636 & 0.7728 & 0.7317  & 0.7461 & 0.5667 & 0.6991 & 0.6863 \\
RoBERTa (from scratch)     & 0.8711 & 0.7445 & 0.8858  & 0.7369 & 0.5285 & 0.5575 & 0.6674 \\ \hline
\multicolumn{8}{c}{RoBERTa models pre-trained by random subsequence masking}\\ \hline
RoBERTa (10M, rand. subseq) & 0.8572 & 0.8253 & 0.9284  & 0.7533 & 0.6111 & 0.7006 & 0.6234 \\
RoBERTa (20M, rand. subseq) & 0.9068 & 0.8135 & 0.9011 & 0.7635 & 0.5799 & 0.7477 & 0.6481 \\
RoBERTa (100M, rand. subseq) & 0.9048 & 0.8248 & 0.9167 & 0.7852 & 0.5860 & 0.7683 & 0.6909 \\ \hline
\multicolumn{8}{c}{MoLFormer and RoBERTa models pre-trained by \ours{}}\\ \hline
\ours{}~(MoLFormer, 10M)          & 0.8980 & 0.8044 & \textbf{0.9669}  & 0.7765 & 0.5811 & 0.7633 & 0.6829 \\
\ours{}~(MoLFormer, 20M)          & 0.8976 & 0.8088 & 0.9436  & 0.7793 & 0.5992 & \underline{0.7801} & 0.7185 \\
\ours{}~(MoLFormer, 100M)         & 0.9055 & 0.8040 & 0.9270  & \underline{0.7893} & 0.5786 & 0.7690 & 0.6017 \\ 
\ours{}~(RoBERTa, 10M)              & 0.8870 & 0.8265 & 0.9258  & 0.7545 & \underline{0.6054} & 0.7106 & 0.6103 \\
\ours{}~(RoBERTa, 20M)              & \textbf{0.9378} & \underline{0.8458} & 0.8919  & 0.7603 & 0.5908 & 0.7594 & 0.6428 \\
\ours{}~(RoBERTa, 100M)             & 0.9237 & 0.7981 & \underline{0.9606}  & \textbf{0.7896} & 0.6042 & \textbf{0.7807} & \textbf{0.7990} \\
\bottomrule
\end{tabular}
}
\label{exp:cls}
\end{table*}

\subsubsection{Classification Tasks}
We choose seven classification tasks from the MoleculeNet benchmark with six baseline models to evaluate and compare the performance of \ours{}. Based on the experimental results shown in Table~\ref{exp:cls}, we can conclude that \ours{}, employing either MoLFormer or RoBERTa architectures, surpasses all of the baselines in five (BBBP, ClinTox, Tox21, HIV, and MUV) out of seven benchmarks and comes a close second in the other two (BACE and SIDER). This result demonstrates the superiority of \ours{} in dealing with the prediction of molecular properties; especially compared within SMILES-based solutions, \ours{} delivers the highest classification accuracy. GEM outperforms \ours{} in BACE and SIDER datasets, which could be attributed to its utilization of explicit 3D structural information of molecules.

\begin{table*}[!t]
\centering
\caption{Comparison of regression errors between fine-tuned \ours{} and existing pre-trained/self-supervised baselines on multiple regression benchmarks.}
\scalebox{1}{
\begin{tabular}{cccccc}
\toprule
& \multicolumn{3}{c}{RMSE} & \multicolumn{2}{c}{MAE} \\ 
\hline
& ESOL        & FreeSolv & Lipo   & QM7      & QM8    \\ \hline
No. molecules & 1,128 & 642 & 4,200 & 6,830 & 21,786 \\
No. prediction tasks & 1 & 1 & 1 & 1 & 12 \\
\hline
\multicolumn{6}{c}{Pre-trained models from existing literature}\\ \hline
MolCLR-gin                  & 1.4717      & 2.7116   & 0.7411 & 96.5469  & 0.0205 \\
MolCLR-gcn                  & 1.5074      & 2.7273   & 0.9033 & 93.5973  & 0.0235 \\
GROVER-base                 & 0.8813      & \underline{1.7772}   & 0.6664 & 101.2853 & 0.0228 \\
GROVER-large                & 0.8831      & 2.7143   & 0.7063 & 114.3004 & 0.0234 \\
GEM                         & 0.7614      & 2.4581   & 0.6861 & \underline{65.0067}  & \underline{0.0179} \\
MoLFormer                   & 0.6613      & 4.4485   & 0.4457 & 69.0700  & \textbf{0.0177} \\ \hline
\multicolumn{6}{c}{MoLFormer and RoBERTa models without pre-training}\\ \hline
MoLFormer (from scratch)  & 0.9721      & 3.3689   & 0.9500 & 68.6214  & 0.0279 \\
RoBERTa  (from scratch)     & 0.9513      & 3.6014   & 0.9910 & 66.8488  & 0.0244 \\ \hline
\multicolumn{6}{c}{RoBERTa models pre-trained by random subsequence masking}\\ \hline
RoBERTa  (10M, rand. subseq) & 0.4909      & 4.4444   & 0.4515 & 68.4687  & 0.0219 \\
RoBERTa  (20M, rand. subseq) & 0.4596 & 3.1672 & 0.4560 & 70.1688 & 0.0207 \\
RoBERTa  (100M, rand. subseq) & 0.4301 & 2.4527 & 0.4430 & 76.6563 & 0.0232\\ \hline
\multicolumn{6}{c}{MoLFormer and RoBERTa models pre-trained by \ours{}}\\ \hline
\ours{}~(MoLFormer, 10M)           & \textbf{0.3432}      & 5.5461   & 0.4919 & 67.7549  & 0.0221 \\
\ours{}~(MoLFormer, 20M)           & 0.4407      & 3.7525   & 0.4325 & 66.2175  & 0.0226 \\
\ours{}~(MoLFormer, 100M)          & 0.5135 & 3.2596   & \underline{0.4272} & 69.3677  & 0.0212 \\ 
\ours{}~(RoBERTa, 10M)               & 0.5707      & \textbf{1.7430}   & 0.4892 & 66.9334  & 0.0212 \\
\ours{}~(RoBERTa, 20M)               & 0.6668      & 1.9143   & 0.6711 & \textbf{64.1665}  & 0.0220 \\
\ours{}~(RoBERTa, 100M)              & \underline{0.3901}      & 3.0487   & \textbf{0.3984} & 74.8177  & 0.0202 \\
\bottomrule
\end{tabular}
}
\label{exp:reg}
\end{table*}

\subsubsection{Regression Tasks}
We choose five classification tasks from the MoleculeNet benchmark with six baseline models to evaluate the performance of \ours{}. Based on the experimental results shown in Table~\ref{exp:reg}, we can conclude that \ours{} exceeds the performance of all baselines in four out of five benchmarks (ESOL, FreeSolv, Lipo, and QM7) and attains comparable results on the qm8 dataset. Especially, \ours{} showcases notable performance gains over the second-best model on ESOL dataset with 41.01\% improvement. In particular, the QM8 dataset involve prediction of several quantum-chemical measures, which is considered challenging without 3D information~\cite{ross2022large}. GEM and MoLFormer lead \ours{} by a slight margin on the QM8 dataset, possibly because GEM incorporates additional 3D information, while MoLFormer is pre-trained with 11 billion molecules (ten times larger than ours).

\subsection{Ablation Study}
To comprehensively evaluate the effectiveness of \ours{}, we conducted ablation studies to dissect the contribution of key treatments, including pre-training strategies, model architectures, and the size of pre-training datasets, to the overall performance. 

The comparison between the functional group-based random masking, random subsequence masking, and training from scratch, underscores the effectiveness of the unique masking approach proposed by \ours{}. Notably, \ours{} demonstrated a significant performance improvement, for instance, achieving error reductions in the ESOL dataset to 0.3432 using functional group-based masking, compared to 0.4909 with random subsequence masking and 0.9721 when trained from scratch. Moreover, we can observe a clear performance improvement from the vanilla MoLFormer to the MoLFormer models pre-trained by \ours{} in the most datasets for both classification and regression tasks. This observation confirms the advantage of the proposed functional group-aware random masking strategy, even with less molecules for pre-training.
In addition, the comparisons between MoLFormer and RoBERTa models highlight a distinct advantage for the more extensive RoBERTa model in the most cases. For instance, in the regression tasks such as FreeSolv, RoBERTa models pre-trained by \ours{} posted superior results (e.g., error of 1.7430 for the RoBERTa 10M) when compared to the \ours{} pre-trained MoLFormer models under similar conditions (error of 5.5461 for MoLFormer 10M).

Expanding the dataset size typically leads to improved performance for \ours{}, as demonstrated across various benchmarks. For example, RoBERTa models pre-trained with \ours{} achieve an accuracy of 0.6103 on the MUV dataset when utilizing 10 million molecules for pre-training. Meanwhile, this accuracy increases to 0.6428 with a dataset size of 20 million molecules and further rises to 0.7990 when leveraging the entire 100 million molecules for pre-training.  Moreover, training with just 10 million molecules on the ESOL dataset yields an error of 0.3432, which represents a significant improvement over MoLFormer models trained from scratch (0.9721). However, on expanding the training set to 20 million and 100 million molecules, we see a performance dip (0.4407 and 0.5135, respectively), indicating possible overfitting or inefficiencies in handling larger datasets without fine-tuning the methodology accordingly~\cite{nakkiran2021deep}. Thus, while increased dataset sizes generally improve the accuracy of \ours{}, the specific nature of the data and the pre-training techniques also play a critical role in extracting the maximal benefit from larger datasets.

\begin{figure*}[!t]
	\centering 
	\includegraphics[width=0.8\linewidth]{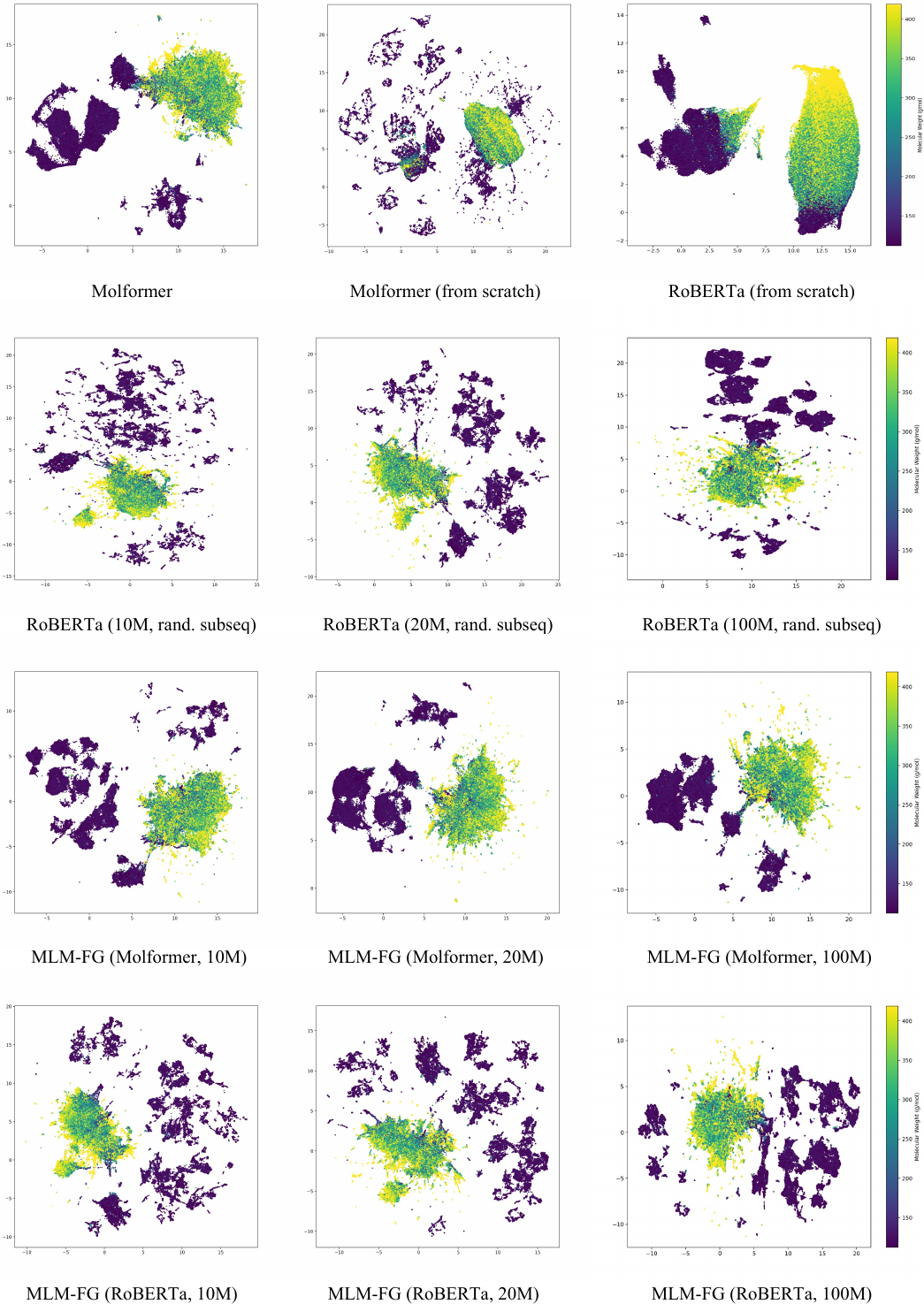} 
	\caption{Visualization of molecular representations learned by \ours{} via UMAP. Representations are extracted from the downstream datasets without finetuned, which contains 312,879 unique molecules. Each point is coloured by its corresponding molecular weight(g/mol).}
    \label{fig:umap}
\end{figure*}

\subsection{Pre-trained Representations Visualization}
The pre-training representation visualization results provide comprehensive insights into the learned molecular representations by \ours{}. 

Our first visualization analysis intends to connect the weights of the molecules and the distribution of their learned representations. These representations, extracted from the downstream datasets without fine-tuning, encompass 312,879 unique molecules. As shown in Figure~\ref{fig:umap}, our experiment maps these presentations onto a 2D space using UMAP~\cite{mcinnes2018umap}, where each point in the visualization is color-coded based on its corresponding molecular weight (g/mol). It can be seen in Figure~\ref{fig:umap} that even without task-specific fine-tuning, \ours{} is capable of distinguishing between light-weighted and heavily-weighted molecules, indicating that the pre-training representation of \ours{} has successfully captured molecular property information. 

\begin{figure*}
	\centering 
	\includegraphics[width=0.6\linewidth]{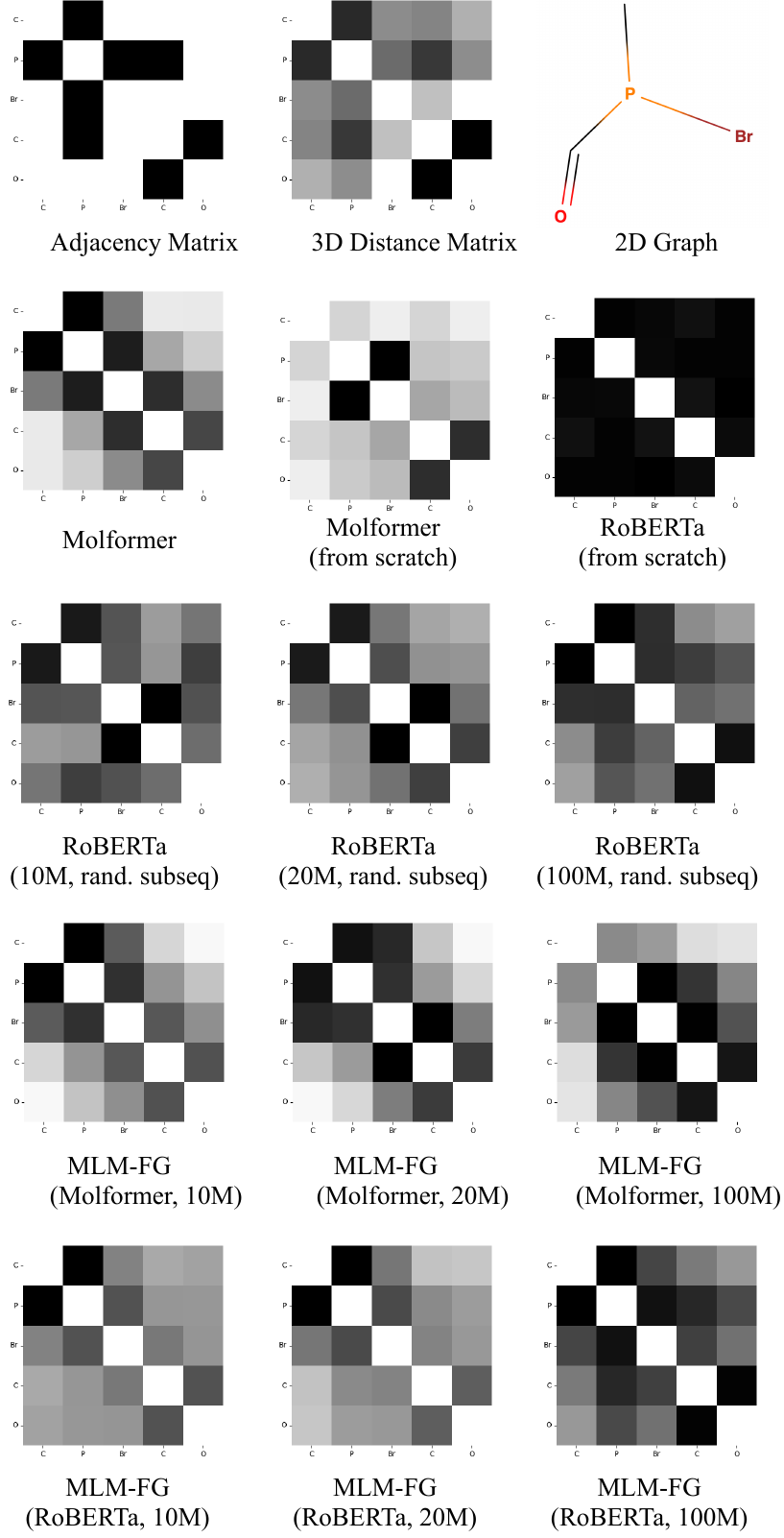} 
	\caption{Visualization of the learned attention map and corresponding molecular structure (bond connectivity and 3D distance in Angstrom) for SMILES ``CP(Br)C=O''.}
    \label{fig:attention}
\end{figure*}

Yet another visualization study has been conducted to analyze the 2D graph and 3D geometric information content in the pre-training representation of \ours{}. This study specifically focused on the comparisons between SMILES-based transformer models, including the standard MoLFormer as well as its variants and RoBERTa pre-trained using \ours{}. For each model, given a SMILES string as the input, we extracted attention vectors for every atomic token, analyzed attentions across different atomic token pairs, and constructed attention matrices for these atoms. We subsequently compare these attention matrices with the corresponding matrices representing covalent bond connectivity and 3D distances between atom pairs. Figure~\ref{fig:attention} showcases the adjacency matrix, 3D distance matrix, and attention matrices for the molecule ``CP(Br)C=O''. It shows that the attention matrices obtained by \ours{} is close to the 3D distance matrix of the molecule, especially when comparing to MoLFormer and MoLFormer/RoBERTa's variants pre-trained by other strategies. 
%
%
Table~\ref{tab:sim_cosine} presents the results for the cosine similarity between the 3D distance matrices and attention matrices, averaged over 50,000 molecules, showcasing the performance of various models. Compared to MoLFormer, MoLFormer trained from scratch, and RoBERTa trained from scratch, \ours{} with both MoLFormer and RoBERTa architectures pre-trained for 100M steps exhibits superior performance in capturing the relationship between attention matrices and 3D distance matrices. 


\begin{table}[htb]
    \caption{Comparisons of cosine similarity between the 3D distance matrix and attention matrix, averaged over 50,000 molecules: The attention matrices of MoLFormer exhibit higher similarity with the 3D distance matrices than those of \ours{} (MoLFormer, 100M). This discrepancy may be attributed to the pre-training of MoLFormer with 1 billion SMILES strings, highlighting the potential impact of pre-training data size on the alignment of attention and 3D geometric information.  The commendable performance of the RoBERTa architecture (for either training from scratch or \ours{}'s variant) underscores the distinct benefits associated with larger-scale architectures.  }
    \centering
    \begin{tabular}{cccccc}
    \toprule
     \thead{MoLFormer} &  \thead{MoLFormer \\ (from scratch)} & \thead{RoBERTa \\ (from scratch)} 
     & \thead{\ours{} \\ (MoLFormer, 100M)} & \thead{\ours{} \\ (RoBERTa, 100M)}  \\
    \midrule
    0.5039 & 0.2174 & 0.8130 
    & 0.3306 & 0.8424  \\
    \bottomrule
    \end{tabular}
    \label{tab:sim_cosine}
\end{table}

\section{Discussion}
This work proposes the \ours{} framework that incorporates functional groups -- as prior information on molecular structures -- and leverages a functional group-aware random masking strategy to pre-train molecular language models on large-scale SMILES databases, yielding enhanced performance and generalization capability for downstream tasks. Our model has been evaluated across 12 datasets, including 7 classification and 5 regression tasks, outperforming the existing state-of-the-art models in 9 of these datasets. Notably, 5 of these datasets involve predictions on multiple sub-tasks of molecular properties, with up to 27 sub-tasks. We also validate the impact of data scale and model type, pre-training on MoLFormer and RoBERTa models across datasets of 10 million, 20 million, and 100 million molecules, followed by an analysis of downstream task performance. We employe AUC, RMSE, and MAE metrics to ensure a fair comparison among molecule analysis methods. Currently, there is limited research on leveraging Transformer architectures with added external knowledge for molecule data analysis. As a publicly available tool, \ours{} offers a powerful resource for molecule analysis and further advanced applications.

\subsection{Key Findings}
Several key findings of this work could be summarized as follows.
\begin{itemize}
    \item  Compared to 2D/3D molecular graphs, SMILES strings lack of explicit structural information and are with limited topological awareness. In the meanwhile, the proposed \ours{} framework overcomes these limitations by incorporating functional group-aware random masking during pre-training, which enables implicit learning of structural features and functional group interactions from SMILES data, ultimately leading to more accurate predictions of molecular properties. 
    
    \item The functional group-aware random masking strategy proposed by \ours{} demonstrates a significant performance improvement compared to masking strategies used in MoLFormer, random subsequence masking, and training from scratch. Furthermore, our analysis indicates that the distances between attention vectors of two atomic tokens extracted from \ours{} closely approximate the actual 3D distances between atoms, providing a more accurate representation of molecular structures.

    \item It has been observed that leveraging a larger model like RoBERTa or pre-training with a larger volume of data typically results in enhanced performance in downstream tasks, especially in the experiments for classification tasks. However, it is important to note that in many scenarios, employing larger models with more data may actually hurt the performance~\cite{nakkiran2021deep}.
 
\end{itemize}

\subsection{Implications}
The \ours{} model represents a significant advancement in molecular modeling by capturing essential structural information through functional group-aware masking within SMILES strings. This capability enhances the prediction of molecular properties and may aid in understanding of structure-activity relationships, making it a valuable tool across drug discovery and metabolomics studies.

In drug discovery, \ours{} can be applied to virtual screening of large compound libraries to identify potential drug candidates, help prioritize compounds that are more likely to interact with specific biological targets, and aid in optimizing design of lead compounds. Additionally, \ours{} may facilitate drug repurposing by screening existing drugs for new therapeutic targets, broadening the utility of known compounds. In metabolomics studies, \ours{} may help identify unknown metabolites, providing valuable insights into metabolic processes and potential therapeutic targets.

Overall, \ours{} emerges as a transformative tool in computational chemistry with broad implications and applications across the life sciences. Integrating \ours{} into various research workflows can accelerate innovation and yield more efficient and targeted outcomes in their respective fields.

\subsection{Limitations}
The \ours{} model, despite its innovations in molecule analysis, confronts several challenges. Firstly, it cannot model very long SMILES sequences. Those exceeding 512 tokens are truncated, potentially leading to loss of information. Secondly, our focus is on molecular modeling, limiting our ability to extend to predicting chemical reactions between molecules and molecular generation. Additionally, the performance of \ours{} could be further improved through incorporating 3D information of molecules in pre-training, fine-tuning, and testing. Moreover, data re-sampling of pre-training datasets and advanced fine-tuning strategies could enhance \ours{} in downstream tasks. Our future work would address these issues for potential performance enhancements.

\section{Methods}
This section provides a comprehensive overview of the design features associated with each component of \ours{}. We will introduce the model architecture of \ours{} and the pre-training strategy.

\subsection{Model Architectures} 
In this work, we present \ours{}, an approach for large-scale pre-training of molecules based on the Transformer blocks, which incorporates multi-layer and multi-head transformer blocks. Specifically, \ours{} offers the same architectural configuration as the one shared by MoLFormer and RoBERTa, employing a 12-layer transformer and a hidden state dimension of $D_h \text{=} 768$. Consider an input SMILES sequence denoted as $\boldsymbol{s}\text{=}(s_1, s_2, ..., s_L)$, where $L$ represents the length of the sequence. \ours{} first tokenizes the sequence and subsequently feeds them into the transformer. This process enables us to extract token embeddings $\boldsymbol{h}\text{=}(h_1, h_2, ..., h_l)\in \mathcal{R}^{L\times D_h}$, where $D_h$ represents the dimension of the hidden representations for the tokens. Then the model takes the series of token embeddings as input and transform them into a lower-dimentional vector to output the embedding of the SMILES sequence. The total number of trainable parameters in \ours{} (MoLFormer) is approximately 48.1M and \ours{} (RoBERTa) is approximately 93.8M. 

\subsection{Pre-training Datasets}
Following many other pre-training based approaches, ours \ours{} is structured into two main phases: pre-training and fine-tuning. During the pre-training phase, which is not tailored to any particular task, \ours{} is trained on a vast corpus of range from 10 million to 100 million SMILES sequences sampled from PubChem~\cite{kim2019pubchem}. The self-supervised training phase enables \ours{} to discern sequential distributions and substructures in molecule sequences, thereby gaining a holistic grasp of their structural and functional insights.

\subsection{Functional group-aware Pre-training Strategy}
Pre-training strategies in molecular representation learning highly correlate with molecule formats. For pre-training with unlabeled data, the prevalent approach involves reconstructing randomly masked tokens in SMILES strings. Given that molecules with similar structures may have vastly different properties, this method might overlook the complex interrelations among molecular features and potentially distort molecular semantics. Our objective is to weave chemical domain knowledge, specifically regarding substructures, into the pre-training process.

Rather than randomly masking subsequences or tokens in SMILES, we mask the cluster of tokens in SMILES that represent these substructures. During the pre-training phase, we start by identifying the substructures that correspond to specific molecular functional groups with RDKit~\cite{fang2022geometry}. Then we randomly selecting a subset of these identified substructures, followed by masking the associated tokens within these substructures. Based on the count of these functional groups within a molecule, our masking strategy adjusts as follows:

\begin{itemize}
\item If a molecule does not contain any functional groups, atom masking is employed as the default strategy. This ensures that the model still learns general structural aspects of molecules lacking specific functional groups.
\item For molecules with fewer than 10 functional groups, we mask only one functional group. This approach is designed to preserve the overall structural integrity of the molecule while still introducing the model to the complexity of functional groups.
\item In cases where a molecule contains more than 10 functional groups, we randomly mask 10\% of these groups. This strategy introduces a higher level of complexity and variability, challenging the model to better generalize its learning across a more diverse set of molecular substructures.
\end{itemize}

The model leverages self-supervised learning to predict masked atoms, thereby acquiring structural information about molecules. This methodical selection and masking process is instrumental in guiding the model to understand and predict the underlying structural characteristics of molecules, enhancing its ability to infer molecular properties and functionalities based on structural cues. By integrating domain knowledge about molecular substructures into our pre-training strategy, we enable the model to develop a more nuanced and accurate representation of molecular structures, paving the way for more effective learning and prediction in downstream tasks.

\subsection{Setups of \ours{} in Experiments}
\ours{} approach gives rise to several model variants distinguished primarily by their underlying architecture and the size of the pre-training dataset. This section introduces the key variants leveraged in our experiments, which include \ours{} (MoLFormer) and \ours{} (RoBERTa).

\begin{itemize}
    \item \emph{\ours{} (MoLFormer)}: This variant utilizes the MoLFormer architecture, specifically designed to capture complex molecular representations using rotary positional embeddings and an efficient linear attention mechanism. The MoLFormer models were pre-trained using three different dataset sizes: 10 million, 20 million, and 100 million molecules. These variations allow for an understanding of how the scale of the pre-training data impacts the effectiveness of model embeddings on downstream tasks, such as molecular property prediction. 

    \item \emph{\ours{} (RoBERTa)}: This variant builds upon the RoBERTa architecture, renowned for its robustness in handling masked language modeling tasks due to its bidirectional encoder representations. Similarly to MoLFormer, RoBERTa models were pre-trained on datasets of 10 million, 20 million, and 100 million molecules. These multiple pre-training data scales enable evaluations of the RoBERTa model's performance adaptability and efficiency when applied to different molecular prediction applications.

\end{itemize}
Both variants of \ours{}, based MoLFormer and RoBERTa transformers, are instrumental in drawing comparative insights between different transformer-based architectures and dataset sizes. The insights gained from these variants help delineate the potential benefits and limitations inherent in each architecture, fostering an advanced understanding of their applicability within molecular informatics.
  
\subsection{Setups of Baseline Methods for Comparisons}
The models we are comparing against are based on cutting-edge methodologies derived from contemporary literature. These methods are as follows. 
\begin{itemize}
    \item \emph{MolCLR-gin and MolCLR-gcn}: These are 2D molecular graph-based models designed to leverage molecular graphs. They are equipped with distinct features focusing on graph-based learning paradigms and trained on 10 million unlabelled molecules. The total number of trainable parameters in MolCLR-gin is approximately 2.2M and in MolCLR-gcn is approximately 0.8M.

    \item \emph{GEM}: A 3D molecular graph-based model, GEM, is built upon a geometry-based approach. It incorporates innovative strategies based on molecular geometry in a 3D space for pre-training on 20 million unlabelled molecules. The total number of trainable parameters in GEM is approximately 0.107M.

    \item \emph{GROVER-base and GROVER-large}: These methods integrate Message Passing Networks into the Transformer-style architecture. They predict contextual properties based on atomic embeddings, encoding contextual information into node embeddings. The dataset for pre-training includes 10 million molecules. The total number of trainable parameters in GROVER-base is approximately 48M and in GROVER-large is approximately 100M.

    \item \emph{MoLFormer}: It is another model based on SMILES representations. This model employs pre-trained representations to capture molecular information encoded as SMILES strings. The pre-training dataset contains 1.1 billion molecules, and the total number of trainable parameters in MoLFormer is approximately 48.1M. 

\end{itemize}
These methods are included for comparison due to their representation of state-of-the-art molecular modeling techniques, each offering distinct advantages. MolCLR-gin and MolCLR-gcn focus on 2D graph representations, GEM provides a 3D approach, GROVER integrates Message Passing Networks with Transformer architectures for contextual analysis, and MoLFormer utilizes SMILES representations with extensive pre-training. Comparing against these varied advanced models allows a comprehensive evaluation of our proposed models' effectiveness and improvements in predictive accuracy.  


\subsection{Hyper-parameters and Training Details}

In the \ours{}, both MoLFormer and RoBERTa comprise 12 layers, each equipped with 12 attention heads. For pre-training, we initialized with a learning rate of 3$\times 10^{-5}$, gradually reducing it using a LambdaLR scheduler, and utilized the AdamW optimizer with a batch size of 1,024 across 16 NVIDIA V100 GPUs. We conducted 50 epochs for datasets of 10M and 20M molecules and reduced the epoch count to 20 for the 100M dataset to balance computational demands and training depth. In the fine-tuning phase, we maintained the learning rate at 3$\times 10^{-5}$ but switched to the FusedLAMB optimizer for better efficiency, with a smaller batch size of 64 to ensure precise model adjustments tailored to specific tasks. 

\section*{Data Availability}
The datasets used for pre-training and fine-tuning are derived from previous studies. These datasets are publicly available via download links as follows. 
\begin{itemize}
    \item PubChem: \url{https://pubchem.ncbi.nlm.nih.gov/} 
    \item QM8: \url{https://moleculenet.org/datasets-1}
    \item ESOL: \url{https://moleculenet.org/datasets-1}
    \item FreeSolv: \url{https://moleculenet.org/datasets-1}
    \item MUV: \url{https://moleculenet.org/datasets-1}
    \item BBBP: \url{https://moleculenet.org/datasets-1} 
    \item BACE: \url{https://moleculenet.org/datasets-1}
    \item ClinTox: \url{https://moleculenet.org/datasets-1}
    \item Tox21: \url{https://moleculenet.org/datasets-1}
    \item SIDER: \url{https://moleculenet.org/datasets-1}
    \item HIV: \url{https://moleculenet.org/datasets-1}
\end{itemize}

\section*{Code Availability}
We built \ours{} using Python and PyTorch. The code repository of \ours{}, readme files and tutorials are all available at \url{https://anonymous.4open.science/r/MLM-FG/README.md}. The checkpoints of pre-trained models are available for download at \url{https://drive.google.com/drive/folders/16vOW0rzMJJAC0iNFbzb6E_40yQ3lbLaF}. 

\section*{Author Contributions}
All authors have made contributions in this paper. T. Peng designed studies conducted experiments and wrote part of the manuscript. Y. Li, X. Li, J. Bian, Z. Xie, N. Sui, S. Mumtaz, and Y. Xu involved in the discussion and wrote part of the manuscript. L. Kong oversaw the research progress, involved in the discussion and wrote part of the manuscript. H. Xiong oversaw the research progress, designed the study and experiments, involved in the discussion, and wrote the manuscript. T. Peng and H. Xiong made the equal technical contributions to this work. H. Xiong is the senior author and L. Kong shares the co-senior contribution.

\section*{Conflict of Interest Statement}
The authors declare that they have no competing interests.

\bibliographystyle{unsrt}
\bibliography{ref}
\end{document}


\maketitle
\section{Framework of \ours{}}

The proposed Molecular Language Model with Random Functional Group Masking (\ours{}) operates within a comprehensive architecture designed to maximize the effective capture of functional group information from SMILES sequences. Key features include:

\begin{enumerate}
    \item \textbf{Model Architecture:} \ours{} utilizes 12-layer multi-head transformer blocks, adapting architectures such as MoLFormer and RoBERTa with a hidden state dimension of $D_h = 768$ for both pre-training and fine-tuning phases. This configuration ensures robust feature extraction capabilities, tailored to the complexities of molecular data represented through SMILES.
    
    \item \textbf{Pre-training Strategy:} The model is pre-trained using a functional group-aware random masking strategy. Specifically, it processes 10 to 100 million SMILES sequences from the PubChem database, introducing random masking aligned with specific molecular functional groups. This approach encourages the model to infer structural properties and interrelationships that are crucial for accurate molecular property prediction.
    
    \item \textbf{Fine-tuning for Diverse Applications:} Once pre-trained, the model supports fine-tuning across a wide array of molecular machine learning tasks. This flexibility underscores \ours{}’s potential applicability across diverse domains such as drug discovery, material science, and computational chemistry.
\end{enumerate}

\section{File Structure}
The repository at \url{https://anonymous.4open.science/r/MLM-FG/README.md} is organized into several directories, each with distinct purposes:
\begin{itemize}
    \item \texttt{\bf data/}: This folder contains datasets for model pre-training and fine-tuning. The datasets include SMILES sequences acquired from PubChem across various scales (10 million, 20 million, 100 million).
    
    \item \texttt{\bf finetuning/}: Contains scripts and resources for fine-tuning the pre-trained models on specific datasets, such as BACE, ClinTox, ESOL, and more. Key scripts like run\_datasets.sh manage the execution of fine-tuning processes.
    
    \item \texttt{\bf pretrain/}: Houses pre-training datasets and vocabularies such as SMILES vocabularies in JSON format, which aid the model's pre-training phase.
    
    \item \texttt{\bf training/}: Involves core scripts and utilities for model pre-training, including shell scripts (run\_RoBERTa\_pretrain.sh, run\_Transformer\_pretrain.sh) that initialize the learning process.
\end{itemize}

\section{Environment Setup}
To utilize the open-source code effectively, the following software dependencies must be installed. 
\subsection{Python Environment Setup}
To successfully replicate and extend upon the experiments performed, set up a dedicated Conda environment:
\begin{verbatim}
conda create --name MolTran_CUDA11 python=3.8.10
conda activate MolTran_CUDA11
\end{verbatim}

\subsection{Conda Install Packages}

Install necessary packages to support the model’s operation:
\begin{verbatim}
conda install pytorch==1.7.1 cudatoolkit=11.0 -c pytorch
conda install numpy=1.22.3 pandas=1.2.4 scikit-learn=0.24.2 scipy=1.6.2
conda install rdkit==2022.03.2 -c conda-forge
\end{verbatim}

\subsection{Pip Install Packages}

Further, use pip to obtain specific packages for model training and visualization:
\begin{verbatim}
pip install transformers==4.6.0 pytorch-lightning==1.1.5
pytorch-fast-transformers==0.4.0 datasets==1.6.2 jupyterlab==3.4.0
ipywidgets==7.7.0 bertviz==1.4.0
\end{verbatim}

\subsection{Apex Compilation}

Due to the employment of Apex optimizers, compile Apex from source for enhanced performance:
\begin{verbatim}
git clone https://github.com/NVIDIA/apex 
cd apex 
git checkout tags/22.03 -b v22.03 
export CUDA_HOME='Cuda 11 install'
pip install -v --disable-pip-version-check --no-cache-dir 
  --no-build-isolation --config-settings "--build-option=--cpp_ext" 
  --config-settings "--build-option=--cuda_ext" ./
\end{verbatim}

\section{Dataset Instructions}
Datasets The datasets used for pre-training and fine-tuning in this study were obtained from previous research and are publicly available. The pre-training dataset was sourced from PubChem, a comprehensive database of chemical molecules and their activities against biological assays. For fine-tuning and evaluation, a diverse set of datasets from MoleculeNet were utilized, including QM8, ESOL, FreeSolv, MUV, BBBP, BACE, ClinTox, Tox21, SIDER, and HIV. These datasets cover a wide range of molecular properties and bioactivities.

The QM8 dataset contains calculated electronic spectra and excited state energy of small molecules. ESOL provides water solubility data for compounds. FreeSolv consists of experimental and calculated hydration free energy of small molecules. MUV, BBBP, BACE, ClinTox, Tox21, SIDER, and HIV are bioactivity datasets that measure the biological effects of molecules against specific protein targets or cellular assays.

All datasets can be accessed via the provided download links in the itemized list. PubChem data is available on the PubChem website, while the MoleculeNet datasets can be downloaded from the MoleculeNet website.
\begin{itemize}
    \item PubChem. \url{https://pubchem.ncbi.nlm.nih.gov/}
    \item MoleculeNet. \url{https://moleculenet.org/datasets-1}
\end{itemize}

\subsection{Pre-training Datasets}

Datasets for the pre-training phase are structured by size (10m, 20m, 100m) and accessible in the designated \texttt{datasets} folder.

\subsection{Fine-tuning Datasets}

Fine-tuning datasets are also provided in the \texttt{datasets} folder, supporting a range of tasks showcased in the paper.

\section{Model Training and Fine-tuning}
The proposed model, \ours{}, employs a combination of pre-training and fine-tuning strategies to effectively learn molecular representations and adapt to specific tasks. The model architecture consists of two main components: MoLFormer and RoBERTa, each comprising 12 layers with 12 attention heads . During the pre-training phase, \ours{} is trained on large-scale datasets containing millions of molecules. The learning rate is initially set to $3\times 10^{-5}$ and gradually reduced using a LambdaLR scheduler. The AdamW optimizer is used with a batch size of 1,024, distributed across 16 NVIDIA V100 GPUs for efficient training. To balance computational demands and training depth, 50 epochs of pre-training are conducted for datasets of 10M and 20M molecules, while the epoch count is reduced to 20 for the larger 100M dataset.

After pre-training, \ours{} undergoes a fine-tuning phase to adapt the learned representations to specific downstream tasks. During fine-tuning, the learning rate is maintained at $3\times 10^{-5}$, but the optimizer is switched to FusedLAMB for improved efficiency. A smaller batch size of 64 is employed to enable precise model adjustments tailored to each task. This fine-tuning strategy allows \ours{} to leverage the knowledge acquired during pre-training and specialize in the target task, resulting in enhanced performance and accuracy. 

\subsection{Pre-training}
Navigate to the \texttt{training/} directory and execute the pre-training scripts, adjusting scales and parameters as needed.
\begin{verbatim}
cd training
sh run_RoBERTa_pretrain.sh
sh run_Transformer_pretrain.sh
\end{verbatim}

\subsection{Fine-tuning}
Switch to the \texttt{finetuning/} directory, where you can refine models using specific datasets.
\begin{verbatim}
cd finetuning
sh run_datasets.sh
\end{verbatim}

\section{Conclusions}
With this organized structure and set-up guidance, our code facilitates robust experimentation and model development in computational chemistry. The \ours{} framework's ability to surpass previous models underscores its potential in molecular representation and prediction tasks. These resources serve as a foundation for future advancements in the field.